\documentclass[11pt,a4paper]{article}
\ifdefined\pdfoutput\pdfoutput=1\fi
\usepackage[a4paper,margin=2.5cm]{geometry}
\usepackage[english]{babel}
\usepackage{microtype}
\usepackage{enumitem}
\setlist[itemize]{label=--}
\usepackage{booktabs,longtable,array}
\usepackage{setspace}
\usepackage{xcolor}
\usepackage[colorlinks=true,linkcolor=blue!50!black,citecolor=blue!50!black,urlcolor=blue!50!black,unicode=true,
  pdftitle={When Agents Need to Consolidate Memories: What RRR Consolidation Explains, and What It Does Not Explain, in Conversation ESAA},pdfauthor={Elzo Brito dos Santos Filho},pdflang={en}]{hyperref}

\title{\textbf{When Agents Need to Consolidate Memories: What RRR Consolidation Explains, and What It Does Not Explain, in Conversation ESAA}}
\author{Elzo Brito dos Santos Filho\\[0.3em]
  \small CEETEPS --- Centro Paula Souza\\ CGTIC}
\date{November 11, 2026}

\begin{document}
\maketitle

\begin{abstract}
Language-model agents can share conversational history across sessions, but persistence does not guarantee retrieval of current decisions and their provenance. This engineering study presents Memory-TCG, a Conversation ESAA extension inspired by \emph{ringing, resonance and renormalization} (RRR), and an exploratory synthetic evaluation comparing chronological retrieval, two simplified baselines (B, recency-aware, and B\ensuremath{^{\prime}}, authority-aware), full RRR, and variants under three output budgets. With 480 facts and 1,024 tokens, full RRR recalled 0.44 of current decisions, versus 0.06 for chronological retrieval and 0.58 for the oracle ceiling; with 4,096 tokens it recalled all of them. B\ensuremath{^{\prime}} and the no-resonance variant matched full RRR in decision recall across all nine scale--budget combinations, as expected when the active decision set and the adapter ordering are the same. The isolated contribution of ringing to decisions was not measured: comparator D1 cuts facts before the graph is built, and the product retains active decisions by rule. Before the output limit, full-RRR conflict F1 was 0.615 at all three scales; after it, at 480 facts and 1,024 tokens, it was 0.000. Reserving 20\% of the budget for conflicts raised that F1 to 0.431 in that scenario, at the cost of lowering decision recall from 0.44 to 0.34, and to 0.930 at 4,096 tokens without decision loss. Fixtures are repetitive, comparators are simplified, and the evaluation was not independently replicated; the results support an engineering analysis, not a claim of selectionist superiority.
\end{abstract}

\section{Introduction}\label{sec:1}
Using heterogeneous agents on a shared task requires continuity across sessions and tools. A decision made in one conversation may be absent from the next context window; an old summary may survive after a decision changes; and a repetition made by another agent may be mistaken for new evidence. Text persistence is necessary to recover history, but selecting what to present remains a distinct problem.

ESAA establishes the separation between agent intentions and state mutations performed by an orchestrator \cite{ref3}. ESAA-Conversational addresses continuity through a local event log and handoff projections \cite{ref1}. Its focus is recording and exposing conversational state across agents. Haber and Tononi \cite{ref2}, in turn, discuss the cost of plasticity in continually learning systems and propose an analogy with memory consolidation during sleep. The authors present as a testable hypothesis that a selectionist approach to consolidation can outperform instructionist alternatives as memory grows and becomes interconnected.

This paper investigates the following question: \textbf{how can RRR-inspired consolidation be integrated into a shared conversational memory while preserving provenance and exposing the limits of automated decisions, and which components explain the gains?} The answer is presented as an engineering study of Memory-TCG, a component of Conversation ESAA. The contributions are: (i) a mapping between the phases proposed in \cite{ref2} and operations implemented over the history of \cite{ref1}; (ii) an architecture with inspectable proposals, explicit promotion, and reconstruction from events; and (iii) an exploratory evaluation with ablations and an authority-aware baseline, delimiting what the results allow and do not allow attributing to each component.

The term \emph{dream} (\emph{sonhar}) designates a consolidation command here. It does not describe subjective experience, consciousness, or biological sleep in an agent.

\section{Background and Related Work}\label{sec:2}
\subsection{Event-Sourced Conversational Continuity}\label{sec:2.1}
In \cite{ref1}, visible turns are captured by adapters and normalized into an append-only history. Projections such as \texttt{handoff.md}, \texttt{state.md}, decisions, and tasks allow an agent to resume work without direct access to another agent's internal session. Mechanical capture is separated from explicit curation of decisions and tasks. The case study in \cite{ref1} reports 570 laboratory events and describes public version 1.1.0; it does not evaluate the Memory-TCG extension presented in this paper.

This model provides a central property: the view presented to an agent can be regenerated from source records. In the examined Conversation ESAA, however, this property has an operational exception: an event-contract repair routine in the PowerShell synchronization path rewrites \texttt{activity.\allowbreak{}jsonl} in full, without a backup copy, when it encounters synchronized events lacking \texttt{agent\_id} or \texttt{workspace\_\allowbreak{}root}, stamping the current \texttt{workspace\_\allowbreak{}root} onto them. Moreover, an append-only history does not by itself resolve the distinction between current and obsolete information. Nor does it guarantee that a token-bounded view will include all relevant evidence.

\subsection{Other Memory Systems}\label{sec:2.2}
MemGPT \cite{ref4} organizes memory into tiers to extend the effective context of models. Generative Agents \cite{ref5} maintains experience records and synthesizes reflections that feed planning. Mem0 \cite{ref6} proposes extraction, update, and retrieval of conversational memories; baselines B and B\ensuremath{^{\prime}} in this work are merely local simplifications inspired by that class of update policies, without executing Mem0. Sleep-time Compute \cite{ref7} shifts model computation ahead of queries; Memory-TCG differs by executing its offline cycle without model calls. Tononi and Cirelli's synaptic homeostasis hypothesis \cite{ref8} provides the biological context for the discussion in \cite{ref2}, without implying functional equivalence between brain and software.

Recent surveys situate \cite{ref1} as a reference for event-sourced persistence in agent memory and evolution architectures \cite{ref10,ref11}. In the multi-agent memory survey by Wang, Zhang et al.\ \cite{ref10}, \cite{ref1} exemplifies collective-state formation and dyadic handoff with structural provenance preservation, within a landscape where compaction frequently erases lineage (\emph{provenance laundering}) and mechanisms for reconciliation, revocation, and recovery remain scarce. In the reliable self-evolution taxonomy by Wang, Hou et al.\ \cite{ref11}, \cite{ref1} is classified at the scaffold level (L2), where reliability requires versioned lineage, a tested rollback path, and matched-budget comparison to prevent \emph{silent store drift}. Memory-TCG extends the scope evaluated in \cite{ref1} along both axes: it introduces non-destructive deprioritization in the active view, candidate conflict detection without automatic merging, watermark-controlled promotion with replay-based rollback, and evaluation under explicit token budgets.

\subsection{The RRR Hypothesis}\label{sec:2.3}
Haber and Tononi \cite{ref2} describe RRR as a selectionist perspective on consolidation: circuits are sampled (\emph{ringing}), their coherence or recurrence dynamics are evaluated (\emph{resonance}), and connections are selectively weakened or preserved (\emph{renormalization}). For digital systems, the authors suggest possibilities such as traversing memory traces outside an active query, evaluating semantic alignment, and reducing redundancies. These suggestions do not constitute a single executable specification.

The paper also predicts that selectionist logic may hold advantages over instructionist strategies, in which an explicit operation updates each memory. That prediction is an empirical hypothesis, not a result demonstrated by the engineering study presented here. The examined implementation uses programmed policies and therefore is not a full realization of the selectionist competition described in \cite{ref2}.

\section{Architecture and Implementation}\label{sec:3}
\subsection{From Conversation to Graph}\label{sec:3.1}
Conversation ESAA maintains \texttt{activity.\allowbreak{}jsonl} as the canonical source of conversational events. The file is designed as append-only, with the operational exception described in Section\nobreakspace{}\ref{sec:2.1}. An adapter creates a workspace snapshot, identifies its event marker (\emph{watermark}), and normalizes records into facts. The Memory-TCG core builds a graph with episodes, decisions, topics, and tasks. Each node retains evidence identifiers; derived relations require an identifiable rule and origin.

The graph distinguishes derived relations from candidate relations. An explicitly recorded supersession can support marking a decision as historical. A lexical association or contradiction suggested by heuristics remains a candidate; its score does not automatically turn it into a verified fact. This separation reduces the risk of converting textual similarity or repetition into authority.

\subsection{Mapping the Three Phases}\label{sec:3.2}
\begin{center}
\begin{tabular}{>{\raggedright\arraybackslash}p{2.9cm}>{\raggedright\arraybackslash}p{5.6cm}>{\raggedright\arraybackslash}p{5.9cm}}
\toprule
\textbf{Phase} & \textbf{Implementation in Memory-TCG} & \textbf{Boundary of Correspondence} \\
\midrule
\emph{Ringing} & Sampling by topic and time window, with reproducible seed and configurable budget; traversal considers graph relations. & The procedure is a deterministic sampling policy, not a reproduction of spontaneous neural activity. \\[2pt]
\emph{Resonance} & Identification of exact duplicates, lexical overlap, candidate relations, and selected structured conflicts. & Lexical similarity does not guarantee semantic equivalence; conflicts outside the rules may go unnoticed. \\[2pt]
\emph{Renormalization} & Proposal for retention, duplicate grouping, deprioritization, historical marking, and conflict flagging. & Selection follows explicit rules; it does not necessarily generate high-level abstractions or demonstrate emergent competition. \\[2pt]
\bottomrule
\end{tabular}
\end{center}
The cycle operates offline, without language-model calls and without mandatory embeddings. Given the same input, policy, and seed, execution is designed to produce reproducible output. Deprioritization acts on the active view; it does not physically remove events from the history. In the evaluated policy, the full graph is built before sampling, and renormalization preserves active decisions even when they are not among the sampled nodes. Workspace identity incorporates its canonical path: the same content in another directory may generate distinct identifiers, even though the metrics coincide.

The numerical results in this paper were obtained with Conversation ESAA v1.3.1, the Memory-TCG 0.1.0 engine, and policy \texttt{rrr-\allowbreak{}policy.\allowbreak{}v1}.

\subsection{Isolation, Promotion, and Attribution}\label{sec:3.3}
In \emph{shadow} mode, consolidation generates a proposal that can be inspected before altering active memory. Upon promotion, the system compares the proposal watermark against the current state; if the history has advanced, a stale proposal is rejected. Acceptance generates an auditable event and materializes the active view. Rollback records another transition, preserving history and enabling reconstruction via replay.

Curated events record \texttt{actor: assistant}, even when triggered by an operator; the field does not prove authorship. Declared identity is carried in \texttt{agent\_id} and classified by \texttt{agent\_\allowbreak{}id\_\allowbreak{}source} as \texttt{declared} or \texttt{unknown}. Missing identity is represented as unknown, and declaration does not authenticate the executor. In historical events, some identities were populated with hardcoded values in the program; therefore, at query time, certain older records are marked \texttt{legacy\_\allowbreak{}unverified} without rewriting history, and should not be used in isolation as proof of authorship. Accepted proposals can be inspected from the canonical event when their \emph{shadow} file is unavailable; a proposal not found in either source remains a documentation gap.

\section{Evaluation Method}\label{sec:4}
\subsection{Scenarios and Arms}\label{sec:4.1}
Synthetic fixtures with 48, 480, and 1,920 facts were used, containing current and superseded decisions, echoes, duplicates, and conflicts. Seeds 11, 29, 42, 73, and 101 alter sampling and do not represent five independent datasets. The evaluation arms are:

\begin{itemize}\setlength{\itemsep}{1pt}
\item \textbf{A} --- chronological window;
\item \textbf{B} --- simplified local policy retaining the most recent event per topic and key, without distinguishing decision from echo;
\item \textbf{B\ensuremath{^{\prime}}} --- simplified local policy retaining the most recent explicit decision per topic and key and preventing an echo from overwriting it; it receives the same structured keys and evidence from the fixtures and does not detect conflicts;
\item \textbf{C} --- full RRR with common ordering;
\item \textbf{CP} --- RRR with 56\% partial sampling;
\item \textbf{CN} --- full RRR reproducing the production \texttt{cmdContext} ordering;
\item \textbf{D1} --- harness comparator that truncates facts by FIFO before building the graph; it is not an isolated replacement of the sampler in the product path;
\item \textbf{D2} --- RRR without \emph{resonance};
\item \textbf{CQ} --- same candidates as C, with 20\% of the budget reserved for conflicts in the canonical proposal order (\texttt{conflict\_id}).
\end{itemize}
The common adapter serializes active decisions first, in lexical identifier order, followed by other active items, and finally candidate conflicts. Because decision ordering is identical across all arms, arms exposing the same set of active decisions coincide in recall under a given budget. This limits what the ties in Section\nobreakspace{}\ref{sec:5} can show: they indicate that the active set is the same, not that the policies are equivalent under other orderings or datasets.

CN extracts the full ordering from the production read path, including non-active items that consume budget, but uses the same standardized JSON serialization as the other arms. It tests production ordering, not exact Markdown output. Conflict candidates include pairs from \texttt{proposal.\allowbreak{}unresolved}, including possible contradictions; false positives are counted. CQ does not alter consolidation: only the packing policy under the budget limit. The implementation does not export a confidence score per conflict; the order used is that of the proposal.

\subsection{Budget and Oracle}\label{sec:4.2}
Each arm runs in a separate Node process. Limits of 512, 1,024, and 4,096 \texttt{cl100k\_base} tokens (\texttt{tiktoken} 0.12.0) are applied. Cost is the sum of tokens of separately encoded JSON records, including line breaks and excluding chat protocol overhead. After an item that does not fit, the adapter continues examining the remaining items.

The oracle knows the current decisions and orders their records, with an empty provenance field, by lowest cost. Because each decision has unit value and costs are additive, it obtains the maximum number of decisions of that format under the budget. The oracle is deterministic and seed-independent. We verified that, in this dataset, no representation emitted for a true decision costs less than the oracle's. The ceiling is conditioned on serialization and the fixture: it is not a ceiling for conflicts, provenance preservation, or combined objectives.

\subsection{Metrics and Reproduction}\label{sec:4.3}
We report decision recall before and after truncation; conflict precision, recall, and F1; and historical contamination, defined as the fraction of ground-truth obsolete records displayed as active. Conflict precision is undefined when no predictions are emitted. Global provenance recall tracks decision recall exactly, because each decision carries its evidence with it, and is therefore omitted from Table\nobreakspace{}\ref{tab:1}.

We report values across the five seeds as median [minimum, maximum]. \texttt{elapsed\_ms}, peak RSS, individual run data, and output cost are included in the artifacts. The scripts were executed twice; on the same workspace path, deterministic fields and output hashes matched; runtime and RSS were excluded from that check.

An additional set of six pairs was run through the real adapter without supplied conflict keys: three contradictory pairs and three negative pairs. It was produced by the same team and is small. The artifact repository \cite{ref9} (commit \texttt{6a2ac1f9e1\allowbreak{}966403603b\allowbreak{}2f972453d1\allowbreak{}2cdbbea3c3}) contains code, synthetic data, commands, and the report \texttt{CONV-\allowbreak{}RRR-\allowbreak{}V13-\allowbreak{}OUTPUT.\allowbreak{}json} (schema \texttt{conv-\allowbreak{}rrr.\allowbreak{}v13.\allowbreak{}v1}), including B\ensuremath{^{\prime}}, the additional metrics, and the CQ quota. The \texttt{memory context} probe and the node counts described in Section\nobreakspace{}\ref{sec:6} are not part of that commit. The evaluation has not been independently replicated.

\begin{quote}\small \textbf{Note.} Conditional provenance is the fraction of expected evidence IDs present within returned true decisions, undefined when no true decision is returned. We also record the fraction of those decisions carrying all of their expected evidence.\end{quote}

\section{Results}\label{sec:5}
\subsection{Separation Between Consolidation and Output}\label{sec:5.1}
Before the output limit, C, CN, B\ensuremath{^{\prime}}, and D2 retained all current decisions across all three scales (recall 1.000 across all seeds); D1 retained 0.40 at 48 facts and 0.56 at 480 and 1,920 facts. Conflict F1 for C and CN before truncation was 0.615 at \textbf{all three scales}, in the exact same proportion: 4 true positives and 5 false positives at 48 facts, 40 and 50 at 480, and 160 and 200 at 1,920 (precision 0.444, recall 1.000). After truncation, C's conflict F1 is 0.000 at the 480 and 1,920 fact scales under 1,024 tokens (Table\nobreakspace{}\ref{tab:1}). Identical post-truncation recall values for C and CP likewise do not establish that they produced the same consolidation: before truncation, CP differs on conflicts (Table\nobreakspace{}\ref{tab:3}). Conditional provenance was 1.000 whenever defined and does not discriminate between arms in this design.

{\footnotesize
\setlength{\tabcolsep}{4pt}
\begin{longtable}{>{\raggedleft\arraybackslash}p{0.9cm}>{\raggedright\arraybackslash}p{1.0cm}>{\raggedright\arraybackslash}p{2.3cm}>{\raggedright\arraybackslash}p{2.3cm}>{\raggedright\arraybackslash}p{2.3cm}>{\raggedright\arraybackslash}p{2.0cm}>{\raggedright\arraybackslash}p{2.3cm}}
\caption{Output under 1,024 tokens (identical values across all five seeds)}\label{tab:1}\\
\toprule
\textbf{Facts} & \textbf{Arm} & \textbf{Decision Recall} & \textbf{Conflict Precision} & \textbf{Conflict Recall} & \textbf{Conflict F1} & \textbf{Historical Contamination} \\
\midrule
\endfirsthead
\toprule
\textbf{Facts} & \textbf{Arm} & \textbf{Decision Recall} & \textbf{Conflict Precision} & \textbf{Conflict Recall} & \textbf{Conflict F1} & \textbf{Historical Contamination} \\
\midrule
\endhead
\bottomrule
\endlastfoot
48 & A & 0.600 & --- & 0.000 & 0.000 & 0.400 \\
48 & B & 0.000 & --- & 0.000 & 0.000 & 0.500 \\
48 & B\ensuremath{^{\prime}} & 1.000 & --- & 0.000 & 0.000 & 0.000 \\
48 & C & 1.000 & 0.444 & 1.000 & 0.615 & 0.000 \\
48 & CP & 1.000 & 0.000 & 0.000 & 0.000 & 0.000 \\
48 & CN & 1.000 & 0.444 & 1.000 & 0.615 & 0.000 \\
48 & D1 & 0.400 & 0.500 & 0.500 & 0.500 & 0.100 \\
48 & D2 & 1.000 & --- & 0.000 & 0.000 & 0.000 \\
\midrule
480 & A & 0.060 & --- & 0.000 & 0.000 & 0.040 \\
480 & B & 0.000 & --- & 0.000 & 0.000 & 0.140 \\
480 & B\ensuremath{^{\prime}} & 0.440 & --- & 0.000 & 0.000 & 0.000 \\
480 & C & 0.440 & --- & 0.000 & 0.000 & 0.000 \\
480 & CP & 0.440 & --- & 0.000 & 0.000 & 0.000 \\
480 & CN & 0.440 & --- & 0.000 & 0.000 & 0.000 \\
480 & D1 & 0.420 & 1.000 & 0.025 & 0.049 & 0.010 \\
480 & D2 & 0.440 & --- & 0.000 & 0.000 & 0.000 \\
\midrule
1920 & A & 0.025 & --- & 0.000 & 0.000 & 0.020 \\
1920 & B & 0.000 & --- & 0.000 & 0.000 & 0.037 \\
1920 & B\ensuremath{^{\prime}} & 0.110 & --- & 0.000 & 0.000 & 0.000 \\
1920 & C & 0.110 & --- & 0.000 & 0.000 & 0.000 \\
1920 & CP & 0.110 & --- & 0.000 & 0.000 & 0.000 \\
1920 & CN & 0.110 & --- & 0.000 & 0.000 & 0.000 \\
1920 & D1 & 0.105 & 1.000 & 0.006 & 0.012 & 0.003 \\
1920 & D2 & 0.110 & --- & 0.000 & 0.000 & 0.000 \\
\end{longtable}
}
{\footnotesize\noindent ``---'': undefined precision (no conflicts returned).\par}

The cross-seed intervals in Table\nobreakspace{}\ref{tab:1} are degenerate: after truncation, all five seeds produced the exact same values. This holds for the displayed metrics and this dataset; the evaluation does not demonstrate that the seed is irrelevant in other fixtures or configurations. Before truncation, conflict detection in the partial arm varied across seeds (Table\nobreakspace{}\ref{tab:3}).

{\small
\begin{table}[htbp]\centering
\setlength{\tabcolsep}{4pt}
\caption{Decision recall by budget and oracle ceiling}\label{tab:2}
\begin{tabular}{>{\raggedleft\arraybackslash}p{1.1cm}>{\raggedleft\arraybackslash}p{1.3cm}>{\raggedleft\arraybackslash}p{2.0cm}>{\raggedleft\arraybackslash}p{2.3cm}>{\raggedleft\arraybackslash}p{1.2cm}>{\raggedleft\arraybackslash}p{2.4cm}>{\raggedleft\arraybackslash}p{3.0cm}}
\toprule
\textbf{Facts} & \textbf{Tokens} & \textbf{C: common order} & \textbf{CN: production order} & \textbf{B\ensuremath{^{\prime}}} & \textbf{D2: no \emph{resonance}} & \textbf{Decision oracle (deterministic)} \\
\midrule
48 & 512 & 1.000 & 1.000 & 1.000 & 1.000 & 1.000 \\
48 & 1024 & 1.000 & 1.000 & 1.000 & 1.000 & 1.000 \\
48 & 4096 & 1.000 & 1.000 & 1.000 & 1.000 & 1.000 \\
\midrule
480 & 512 & 0.220 & 0.220 & 0.220 & 0.220 & 0.280 \\
480 & 1024 & 0.440 & 0.440 & 0.440 & 0.440 & 0.580 \\
480 & 4096 & 1.000 & 1.000 & 1.000 & 1.000 & 1.000 \\
\midrule
1920 & 512 & 0.055 & 0.055 & 0.055 & 0.055 & 0.070 \\
1920 & 1024 & 0.110 & 0.110 & 0.110 & 0.110 & 0.145 \\
1920 & 4096 & 0.445 & 0.445 & 0.445 & 0.445 & 0.585 \\
\bottomrule
\end{tabular}
\end{table}
}
At 480 facts and 1,024 tokens, C and CN recalled 0.44 of decisions, versus 0.06 in A: roughly 7.3 times the chronological baseline in this fixture, and approximately 76\% of the conditioned oracle ceiling (0.58). The gap to the oracle does not indicate a retention failure, as all decisions were retained before truncation; it reflects the adapter's lexical ordering, which does not prioritize lower-cost decisions, and the fact that the oracle emits records with empty provenance. At 4,096 tokens and 480 facts, C, CN, B\ensuremath{^{\prime}}, and D2 recalled all decisions, matching the oracle. D2 matches C in decision recall across all nine combinations (Table\nobreakspace{}\ref{tab:2}). The oracle neither preserves evidence nor evaluates conflicts.

{\small
\begin{table}[htbp]\centering
\setlength{\tabcolsep}{4pt}
\caption{Conflict F1 for CP before truncation, by seed}\label{tab:3}
\begin{tabular}{rrrrrrr}
\toprule
\textbf{Facts} & \textbf{Seed 11} & \textbf{Seed 29} & \textbf{Seed 42} & \textbf{Seed 73} & \textbf{Seed 101} & \textbf{Median} \\
\midrule
48 & 0.000 & 0.000 & 0.000 & 0.000 & 0.000 & 0.000 \\
\midrule
480 & 0.033 & 0.000 & 0.152 & 0.033 & 0.031 & 0.033 \\
\midrule
1920 & 0.047 & 0.023 & 0.081 & 0.008 & 0.039 & 0.039 \\
\bottomrule
\end{tabular}
\end{table}
}
In CP, pre-truncation decision recall was 1.000 across all seeds; only conflict detection varied.

\subsection{What the Ablations and B\texorpdfstring{\ensuremath{^{\prime}}}{'} Allow Attributing}\label{sec:5.2}
Variant D2 isolates the absence of \emph{resonance} in this design. D1 applies a cutoff prior to graph construction and therefore does not isolate the product's \emph{ringing} stage.

\begin{itemize}\setlength{\itemsep}{1pt}
\item D2 (no \emph{resonance}) matches C in \textbf{decision recall} across all nine scale and budget combinations (Table\nobreakspace{}\ref{tab:2}). This coincidence does not extend to conflict F1, which is 0.000 in D2.
\item D1 applies FIFO before building the graph and retains, before output truncation, 0.40 of decisions at 48 facts and 0.56 at 480 and 1,920, versus 1.000 in C. After truncation, the difference is 0.60 at 48 facts (all budgets) and 0.44 at 480 facts and 4,096 tokens; in the remaining combinations at 480 and 1,920 facts, it does not exceed 0.03. Because facts discarded by D1 never enter the graph, these differences do not identify the isolated contribution of the product's sampler.
\item B\ensuremath{^{\prime}}, which executes none of the three phases and merely prevents an echo from overwriting an explicit decision, matches C across all nine combinations in Table\nobreakspace{}\ref{tab:2}.
\end{itemize}
The coincidences among C, CN, B\ensuremath{^{\prime}}, and D2 must be read with the caveat from Section\nobreakspace{}\ref{sec:4.1}: the adapter orders decisions identically across all arms, and a recall tie under that ordering shows that the set of active decisions is the same, not that the policies are equivalent. What the data support is narrower: in these fixtures, with explicit keys, an authority-aware rule (B\ensuremath{^{\prime}}) produces the same active decision set as full RRR. The product retains active decisions in the graph by rule, and D1 does not measure the isolated contribution of \emph{ringing}. D2 has a conflict F1 of 0.000 before and after truncation, whereas C reaches 0.615 before truncation; this difference is specific to the tested design and metrics.

Table\nobreakspace{}\ref{tab:3} provides an indication regarding \emph{ringing} that does not pass through decisions. CP, which differs from C only in the sampled fraction (56\%), kept pre-truncation decision recall at 1.000, but reduced conflict F1 from 0.615 (C) to 0.000 at 48 facts and to medians of 0.033 at 480 facts and 0.039 at 1,920 facts. This is consistent with an effect of sampling on conflict and episode detection rather than on decisions, which the product retains by rule. The result was not isolated by an ablation in the product path and remains a hypothesis.

\subsection{Conflicts Under Budget}\label{sec:5.3}
{\footnotesize
\setlength{\tabcolsep}{3pt}
\begin{longtable}{>{\raggedleft\arraybackslash}p{1.0cm}>{\raggedleft\arraybackslash}p{1.2cm}>{\raggedright\arraybackslash}p{1.1cm}>{\raggedleft\arraybackslash}p{1.4cm}>{\raggedleft\arraybackslash}p{1.75cm}>{\raggedleft\arraybackslash}p{0.9cm}>{\raggedleft\arraybackslash}p{0.7cm}>{\raggedleft\arraybackslash}p{0.7cm}>{\raggedleft\arraybackslash}p{0.7cm}>{\raggedleft\arraybackslash}p{1.35cm}>{\raggedleft\arraybackslash}p{1.7cm}}
\caption{Conflict precision, recall, and F1 by budget (identical values across all five seeds)}\label{tab:4}\\
\toprule
\textbf{Facts} & \textbf{Tokens} & \textbf{Arm} & \textbf{Precision} & \textbf{Recall} & \textbf{F1} & \textbf{TP} & \textbf{FP} & \textbf{FN} & \textbf{Tokens Used} & \textbf{Decision Recall} \\
\midrule
\endfirsthead
\toprule
\textbf{Facts} & \textbf{Tokens} & \textbf{Arm} & \textbf{Precision} & \textbf{Recall} & \textbf{F1} & \textbf{TP} & \textbf{FP} & \textbf{FN} & \textbf{Tokens Used} & \textbf{Decision Recall} \\
\midrule
\endhead
\bottomrule
\endlastfoot
48 & 512 & C & 0.800 & 1.000 & 0.889 & 4 & 1 & 0 & 512 & 1.000 \\
48 & 1024 & C & 0.444 & 1.000 & 0.615 & 4 & 5 & 0 & 572 & 1.000 \\
48 & 4096 & C & 0.444 & 1.000 & 0.615 & 4 & 5 & 0 & 572 & 1.000 \\
48 & 512 & CN & 0.375 & 0.750 & 0.500 & 3 & 5 & 1 & 504 & 1.000 \\
48 & 1024 & CN & 0.444 & 1.000 & 0.615 & 4 & 5 & 0 & 892 & 1.000 \\
48 & 4096 & CN & 0.444 & 1.000 & 0.615 & 4 & 5 & 0 & 892 & 1.000 \\
48 & 512 & D1 & 0.500 & 0.500 & 0.500 & 2 & 2 & 2 & 276 & 0.400 \\
48 & 1024 & D1 & 0.500 & 0.500 & 0.500 & 2 & 2 & 2 & 276 & 0.400 \\
48 & 4096 & D1 & 0.500 & 0.500 & 0.500 & 2 & 2 & 2 & 276 & 0.400 \\
48 & 512 & CQ & 0.571 & 1.000 & 0.727 & 4 & 3 & 0 & 503 & 1.000 \\
48 & 1024 & CQ & 0.444 & 1.000 & 0.615 & 4 & 5 & 0 & 572 & 1.000 \\
48 & 4096 & CQ & 0.444 & 1.000 & 0.615 & 4 & 5 & 0 & 572 & 1.000 \\
\midrule
480 & 512 & C & --- & 0.000 & 0.000 & 0 & 0 & 40 & 506 & 0.220 \\
480 & 1024 & C & --- & 0.000 & 0.000 & 0 & 0 & 40 & 1,012 & 0.440 \\
480 & 4096 & C & --- & 0.000 & 0.000 & 0 & 0 & 40 & 4,094 & 1.000 \\
480 & 512 & CN & --- & 0.000 & 0.000 & 0 & 0 & 40 & 506 & 0.220 \\
480 & 1024 & CN & --- & 0.000 & 0.000 & 0 & 0 & 40 & 1,012 & 0.440 \\
480 & 4096 & CN & --- & 0.000 & 0.000 & 0 & 0 & 40 & 4,086 & 1.000 \\
480 & 512 & D1 & --- & 0.000 & 0.000 & 0 & 0 & 40 & 500 & 0.200 \\
480 & 1024 & D1 & 1.000 & 0.025 & 0.049 & 1 & 0 & 39 & 1,024 & 0.420 \\
480 & 4096 & D1 & 0.468 & 0.550 & 0.506 & 22 & 25 & 18 & 4,089 & 0.560 \\
480 & 512 & CQ & 1.000 & 0.125 & 0.222 & 5 & 0 & 35 & 504 & 0.180 \\
480 & 1024 & CQ & 1.000 & 0.275 & 0.431 & 11 & 0 & 29 & 1,019 & 0.340 \\
480 & 4096 & CQ & 0.870 & 1.000 & 0.930 & 40 & 6 & 0 & 4,085 & 1.000 \\
\midrule
1920 & 512 & C & --- & 0.000 & 0.000 & 0 & 0 & 160 & 506 & 0.055 \\
1920 & 1024 & C & --- & 0.000 & 0.000 & 0 & 0 & 160 & 1,012 & 0.110 \\
1920 & 4096 & C & --- & 0.000 & 0.000 & 0 & 0 & 160 & 4,094 & 0.445 \\
1920 & 512 & CN & --- & 0.000 & 0.000 & 0 & 0 & 160 & 506 & 0.055 \\
1920 & 1024 & CN & --- & 0.000 & 0.000 & 0 & 0 & 160 & 1,012 & 0.110 \\
1920 & 4096 & CN & --- & 0.000 & 0.000 & 0 & 0 & 160 & 4,094 & 0.445 \\
1920 & 512 & D1 & --- & 0.000 & 0.000 & 0 & 0 & 160 & 506 & 0.055 \\
1920 & 1024 & D1 & 1.000 & 0.006 & 0.012 & 1 & 0 & 159 & 1,024 & 0.105 \\
1920 & 4096 & D1 & --- & 0.000 & 0.000 & 0 & 0 & 160 & 4,093 & 0.415 \\
1920 & 512 & CQ & 1.000 & 0.031 & 0.061 & 5 & 0 & 155 & 504 & 0.045 \\
1920 & 1024 & CQ & 1.000 & 0.069 & 0.129 & 11 & 0 & 149 & 1,019 & 0.085 \\
1920 & 4096 & CQ & 1.000 & 0.288 & 0.447 & 46 & 0 & 114 & 4,094 & 0.355 \\
\end{longtable}
}
{\footnotesize\noindent ``---'': undefined precision (no conflicts returned). CQ applies a 20\% conflict quota to C.\par}

\textbf{F1 drops as the budget increases when the entire conflict queue fits.} At 48 facts, C retains, under 512 tokens, 4 true conflicts and 1 false positive (precision 0.800, recall 1.000, F1 0.889); under 1,024 tokens, it retains the same 4 true conflicts and 5 false positives (precision 0.444, recall 1.000, F1 0.615). The larger budget adds no true conflicts; it merely admits \texttt{unresolved} candidates that are false positives. In this scenario, the initial segment of the queue contains the four true conflicts and one false positive; this observation does not establish a universal ordering.

\textbf{Conflicts vanish after truncation at larger scales under an ordering that places them last.} At 480 facts, C keeps 50 active decisions and 72 active episodes; under 4,096 tokens, those items consume 4,094 tokens and no conflicts fit, even though decision recall is 1.000 (Table\nobreakspace{}\ref{tab:4}). D1, whose active view is smaller (28 decisions and 50 episodes), leaves room and reaches an F1 of 0.506 (precision 0.468, recall 0.550) in the same scenario. D1's advantage in post-truncation conflicts is an effect of ordering and budget rather than better detection: before truncation, D1's F1 (0.500) is lower than C's (0.615).

\textbf{The 20\% quota (CQ) recovers conflicts without altering consolidation.} At 480 facts and 4,096 tokens, CQ maintains decision recall at 1.000 and raises conflict F1 from 0.000 to 0.930 (40 true positives and 6 false positives, precision 0.870): the cut falls on episodes rather than decisions, because active-item serialization already places decisions first. At 1,024 tokens, F1 rises from 0.000 to 0.431 (11 true positives and 0 false positives, precision 1.000), while decision recall drops from 0.440 to 0.340. At 512 tokens, the pattern repeats on a smaller scale (F1 0.222, precision 1.000; decision recall 0.180 versus 0.220). At 1,920 facts and 4,096 tokens, CQ achieves an F1 of 0.447 (46 true positives and 0 false positives) and decision recall of 0.355, versus 0.445 for C. Overall, the quota reduces decision recall in five of the nine scale--budget combinations. It trades episodes---and, under tight budgets, some decisions---for conflicts located at the head of the queue. At 48 facts and 512 tokens, CQ performs worse than C (F1 0.727 versus 0.889), because the budget already accommodated all true conflicts and the quota brings in false positives earlier.

\textbf{C and CN.} Both orderings produce the exact same decision recall across all scenarios because both serialize active decisions first; the non-active items added by CN enter after them. The difference appears only in what follows: at 48 facts and 512 tokens, CN spends budget on topics and subjects before conflicts and retains 3 true positives and 5 false positives (F1 0.500), versus 0.889 in C. At 1,024 tokens, the budget fits everything and both reach 0.615. Equality in decision count does not imply that the exact same decisions are chosen in all cases.

\subsection{Why B Returns Zero}\label{sec:5.4}
In each topic of the scaled fixture, B retains the most recent event for each key. The subsequent episodic echo overwrites the authoritative decision: consequently, B returns zero current decisions \textbf{even before the output limit}, and its post-limit historical contamination reaches 0.140 at 480 facts and 1,024 tokens. This behavior stems from the baseline's design, which favors recency; B\ensuremath{^{\prime}} corrects this exact point and matches C.

\subsection{Adversarial Pairs}\label{sec:5.5}
All five seeds produced identical results on the six pairs without supplied conflict keys: one true positive, two false negatives, three true negatives, and zero false positives. Conflict precision, recall, and F1 were 1.000, 0.333, and 0.500, respectively. The deadline and TLS conflicts went undetected. This small set, produced by the same team, serves solely to illustrate concrete failure modes \cite{ref9}.

\subsection{Functional Evidence}\label{sec:5.6}
Automated tests in Node exercised \emph{shadow} mode, stale-proposal rejection, promotion, rollback, replay, proposal recovery from the ledger, topic queries, baseline B\ensuremath{^{\prime}}, and isolated installation. The PowerShell layer connecting memory commands to the ledger lacks automated tests for dreaming, acceptance, and rollback; the properties listed above therefore come from the engine and installer tests. These tests do not prove global installation updates or independent review.

\section{Discussion and Threats to Validity}\label{sec:6}
Memory-TCG shows that it is possible to integrate the three named stages of RRR into a shared conversational memory without replacing the source history. The snapshot and watermark allow evaluating a stable version of memory; explicit promotion makes changes to the active view auditable; and the separation between verified and candidate edges prevents a lexical suggestion from automatically becoming an authoritative decision.

\textbf{What explains the results.} Decision recall under budget in this experiment is consistent with \textbf{authority filtering} in the active view when keys are explicit: B\ensuremath{^{\prime}} achieves the same recall as full RRR across all scales. Because the adapter orders decisions identically for all arms, this result shows that the active decision set coincides, not that the policies are equivalent under other orderings. D1's FIFO comparator excludes facts prior to graph construction, whereas the product path preserves active decisions by rule; the isolated contribution of \emph{ringing} to decisions was not measured. Removing \emph{resonance} did not alter decision recall, but it altered pre-truncation conflict detection; at larger scales, that difference does not appear after truncation when conflicts are serialized after active items.

\textbf{What remains for \emph{ringing}.} Because the product retains active decisions by rule, the open question regarding \emph{ringing} concerns conflicts and episodes rather than decisions. Table\nobreakspace{}\ref{tab:3} is consistent with an effect of the sampled fraction on conflict detection, but does not isolate it. The missing ablation is swapping only the sampler in the product path while preserving the full graph, and measuring conflicts and episodes.

\textbf{Conflict quota.} At 480 facts and 4,096 tokens, the 20\% quota keeps decision recall at 1.000 and raises conflict F1 from 0.000 to 0.930 (Table\nobreakspace{}\ref{tab:4}). In the 480-fact scenarios with 512 or 1,024 tokens and in the 1,920-fact scenarios, it recovers conflicts with 1.000 precision while reducing decision recall; that precision does not generalize to the 48-fact scenarios. The ordering used is that of the proposal (\texttt{conflict\_id}), not an exported confidence score; at 48 facts and 512 tokens, precision was 0.800 in C and 0.571 in CQ, showing that the head of the queue is not always optimal. An explicit confidence field remains untested.

\textbf{The prediction of \cite{ref2} was not tested.} The coincidence between B\ensuremath{^{\prime}} and C neither refutes nor confirms the selectionist hypothesis: the fixtures supply structured keys and explicit supersession relations, a condition under which an instructionist rule suffices by construction. The advantage predicted in \cite{ref2} concerns large, interconnected memories lacking explicit keys, which this design does not represent. Furthermore, the evaluated implementation is deterministic and rule-based, not a selectionist competition.

\textbf{Partial correspondence with \cite{ref2}.} Reproducible sampling replaces open-ended stochastic exploration; resonance uses rules and lexical overlap rather than general semantic evaluation; and renormalization is a policy-determined selection. These choices favor inspection and replay, but do not demonstrate that relevance emerges from competition among memories. Nor does the study measure model weight updates or continual parameter learning.

\textbf{Threats to validity.} The scaled fixtures repeat patterns and supply structured conflict keys, which likely explains the identical pre-truncation F1 across scales and the degenerate cross-seed intervals; the keyless set contains only six pairs. By ordering decisions by identifier, the common adapter produces ties across arms that do not generalize to other orderings. The token limit is exact for the chosen tokenizer, but other models may use different tokenizers and context windows. The oracle computes a ceiling solely for decisions, without provenance cost. B and B\ensuremath{^{\prime}} are simplified local baselines; neither represents Mem0 or full instructionist systems. The five seeds must not be treated as five independent samples of real tasks. Conditional provenance was 1.000 whenever defined and does not discriminate between arms in this design. CQ uses the canonical proposal order rather than a learned score. The same team designed, executed, and reviewed the evaluation.

The cross-agent continuity case is observational and is reported in Appendix\nobreakspace{}\ref{sec:A}. Authorship of an older promotion is unverifiable, and the three-identity test is synthetic. Real data and conversational excerpts require human privacy review prior to release.

\textbf{Scale and policy regimes.} Policy scale thresholds alter the executed code path. In the fixtures, each fact becomes an episode or decision node; topics and tasks do not enter these counts. The 48-fact scale produces 48 nodes (38 episodes and 10 decisions) and does not cross the 50-node threshold, the 200-node threshold, or the 250-episode lexical window. The 480-fact scale produces 480 nodes (380 episodes and 100 decisions), crossing the 50- and 200-node thresholds and the 250-episode window. The 1,920-fact scale produces 1,920 nodes (1,520 episodes and 400 decisions) and additionally triggers compaction above 500 nodes, without reducing the graph in this fixture. Comparisons across scales therefore mix size and regime changes.

\textbf{Product output in a synthetic probe.} An additional 480-fact synthetic probe built an active view with the adapter and queried \texttt{memory context}, without real acceptance and without \texttt{pack}. Under a requested budget of 3,200 characters, the output displayed 4 of the 50 current decisions (0.08) and 4 conflicts, while 295 items were omitted. The final Markdown totaled 3,235 characters, including the omission notice---35 characters over the requested budget. This probe is not comparable to arm CN: it uses characters and Markdown, whereas CN uses tokens and standardized JSON serialization, and its numbers are not direct measurements of production Markdown. The probe and the node counts above are not part of the artifacts in \cite{ref9}.

Future evaluations should include independent and more diverse datasets, conflicts expressed without keys, interconnected memories without explicit supersession relations---the regime in which \cite{ref2} predicts a selectionist advantage---, an ablation of the sampler in the product path, a per-conflict confidence score, measurement of production Markdown output, and properly configured external baselines. Independent replication and evaluation on real tasks are required to support broader conclusions. The absence of a conflict marker or a successful structural check does not prove that the entire memory is free of contradictions.

\section{Conclusion}\label{sec:7}
Conversation ESAA provides the foundation for continuity and traceability; Memory-TCG adds offline consolidation inspired by the three RRR phases. Their combination makes it possible to generate, inspect, accept, and roll back an active view without erasing source events. It represents an operational implementation of an \textbf{engineering mapping} between the ideas of both papers, with explicit differences relative to the selectionist hypothesis.

In the evaluated scenarios, full RRR retained all current decisions before truncation and, under 1,024 tokens at 480 facts, recalled 0.44 of them---roughly 7.3 times the chronological window---, reaching full recall at 4,096 tokens. B\ensuremath{^{\prime}} and the no-resonance variant matched full RRR in decision recall, which, under the adapter's common ordering, indicates that the active decision set is the same; it does not indicate equivalence among the policies. The isolated contribution of \emph{ringing} to decisions was not measured, and the product retains active decisions by rule. \emph{Resonance} improves pre-truncation conflict detection, but under an ordering that places conflicts last, they are discarded by the budget at larger scales. Reserving 20\% of the budget for the conflict queue recovers those pairs---at 480 facts and 4,096 tokens, without losing decisions; in five of the nine combinations, the quota reduces decision recall. In a synthetic probe of \texttt{memory context} with a 3,200-character budget, the product output displayed 4 of the 50 current decisions, suggesting that measuring production Markdown output is the necessary next step. Adversarial cases revealed undetected contradictions. These findings justify investigating memory consolidation as a component of multi-agent systems and, specifically, budget-constrained ordering, without concluding that Haber and Tononi's hypothesis has been tested or proven, or that agents experience dreaming.

\appendix
\section{Observed Continuity}\label{sec:A}
In real usage, a handoff decision was recorded, an accepted generation retained it, and a second agent reported having read both the decision and the generation. The evidence shows a sharing sequence, but the authorship field of the historical promotion does not allow identifying its executor with certainty. In another curated topic, three events were explicitly linked; the previously consulted active generation preceded those links. An additional isolated test reproduced handoff across three declared identities. That test validates the workflow on synthetic data; it does not prove execution controlled by three distinct real models.

\end{document}